\documentclass[12pt]{article}
\usepackage{amssymb,amsfonts}
\usepackage[T2A]{fontenc}
\usepackage[cp866]{inputenc}
\inputencoding{cp866}
\begin{document}

\hfil{{\bf ON CANONICAL VARIABLES IN INTEGRABLE}}\vskip0.6cm
  \hfil {\bf {MODELS OF MAGNETS}}\hfil
\vskip0.8cm \hfil{\bf E.SH.GUTSHABASH}\hfil \vskip0.2cm
\hfil{Institute Research for Physics, St.-Petersburg State
University, Russia,}\hfil

\hfil{e-mail: gutshab@EG2097.spb.edu}\hfil \vskip0.8cm \vskip0.8cm

 \hskip4.8cm  {\bf 1. INTRODUCTION} \vskip0.4cm

The most general formulation of phenomenological models of magnets
(or spin systems) which includes all the known completely
integrable ones, has the following form:

$$
{\bf S}_t={\bf F}_0(x, y, {\bf S}, {\bf S}_x, {\bf S}_y, {\bf
S}_{xx}, {\bf S}_{yy}, {\bf S}_{xy}, {\bf J}, u, u_x, u_y, u_{xy},
\alpha^2),
$$
$$
\eqno(1.1) $$
$$
 u_{xx}+\alpha^2u_{yy}=R_0({\bf S}, {\bf S}_x, {\bf
S}_y), $$ where ${\bf S}={\bf S}(x, y, t)$ is the magnetization
vector, ${\bf F}_0(,)$ is a vector-function, $u=u(x,y)$ is an
auxiliary field, $R_0(,)$ is a scalar function, ${\bf J}$ is a set
of constants characterizing the magnet, and $\alpha^2=\pm 1$.

The function ${\bf F}_0$ usually takes the form:

$$ {\bf F}_0={\bf S}\wedge \frac {\:\:\:\delta F_{eff}}{\delta
{\bf S}}+{\bf F}_1,   \eqno(1.2) $$ where $F_{eff}$ is the
functional of the crystal's free energy (throughout the paper the
symbol $\delta/\delta$ stands for the variational derivative). The
first term in the right hand side was suggested by Landau and
Lifshits [1] to describe the exchange interactions.

The representation (1.1)-(1.2) is often inconvenient for solving
problems. One would like to deal with more tractable forms of the
equations (1.1), which, in turn, requires introduction of new
dependent variables. Apparently, such a variable, the
stereographic projection, has been used for the first time in
paper [2] to describe the instanton solutions in the
two-dimensional $O(3)$ $ \sigma $-model (the $ 2D $ stationary
Heisenberg ferromagnet). Later it was exploited in various
situations, see, e. g. [3-5].

In the present paper we show on examples of three models - the
deformed Heisenberg, the Landau - Lifshits, and the Ishimori
magnets - that it is helpful to introduce the corresponding
canonical variables. In particular, they allow to simplify
significantly certain calculations, as compared to the usage of
the ${\bf S}$ variable, and, more substantially, to clarify a set
of questions important both from physical and mathematical
viewpoints. Another argument in their favor is that in these
variables the models fit in a class of models admitting a
differential-geometric interpretation intensively studied recently
[6].

The model of deformed Heisenberg magnet was suggested in [7] where
also an exact solution of it for the case of trivial background
was obtained by the inverse scattering method, and the
conservation laws were calculated. In doing so, it was shown that
perturbations localized in the space are spreaded, that is, the
solutions are instable. The gauge equivalence of this model and
the nonlinear Schr\"odinger equation with an integral nonlinearity
was established in [8] and [9]. The matrix Darboux transform
method was applied in the paper [10], where exact solutions of the
model where calculated on the background of new spiral-logarithmic
structures.

The Landau - Lifshits equation is a subject of vast studies. In
particular, the Lax representation and conservation laws for it in
the completely anisotropic case have first been obtained in [11],
soliton solutions were found by the dressing method in [12].
Hamilton aspects of the equation were analyzed in detail in recent
paper [13].

The Ishimori magnet was also considered in many papers. In
particular, series of exact solutions were obtained in [14] by the
inverse scattering and ${\bar \partial}$ - dressing methods, the
Darboux transform was applied to it in [15] and [16] (in [16] - on
the background of spiral structures). Notice also the important
paper [5], where the gauge equivalence of the Ishimori-II and
Davy-Stewartson-II models was established.

The structure of this paper is as follows. In section 2 we
consider the deformed Heisenberg magnet model, define the
canonical variables and analyze stability of the solutions. In
section 3 the Landau - Lifshits equation is obtained in terms of
the stereographic projection and a stationary version of this
equation is studied. Finally, in section 4 we define two pairs of
canonical variables for the Ishimori model, re-write the model and
the Hamiltonian in these variables, and calculate the Hamiltonian
on some of the simplest known solutions. This is preceded by a
discussion of the physical interpretation of the model. The
Appendix contains a Lax pair for an "extended" \enskip system of
the deformed Heisenberg magnet model.

 \vskip0.8cm \hskip1.0cm  {\bf 2. DEFORMED HEISENBERG MAGNET
 EQUATION}
\vskip0.6cm a). {\em Canonical variables.}

Let us consider the deformed Heisenberg magnet equation
[7]{\footnote {This equation can be thought of as a
cylindrical-symmetric reduction of the (2+1)-dimensional
non-integrable Landau - Lifshits equation, $ {\bf S}_t={\bf
S}\wedge \triangle {\bf S}.$ The relation between the latter and
the system of coupled nonlinear Schr\"odinger equations in the
dimension (2+1) has been discussed in detail in [17].}}:

$$ {\bf S}_t={\bf S}\wedge {\bf S}_{xx}+ \frac{1}{x}{\bf S}\wedge
{\bf S}_x . \eqno(2.1) $$ Here $x=\sqrt{x_1^2+x_2^2} >
0,\:x_1,\:x_2$ are the Cartesian coordinates on the plane, ${\bf
S}(x,t)=(S_1, S_2, S_3),\:|{\bf S}|=1$.

The phase space for this equation is generated by initial data
$(S_1,\:S_2,\:S_3)$ subject to the constraint $|{\bf S}|=1$. The
Poisson brackets of the canonical variables $S_i$ in the model
satisfy the standard relations:

$$
\{S_i(x), S_j(y) \}=-\varepsilon_{ijk}S_k(x)\delta(x-y), \:\:\:i,
j, k=1,2.3, \eqno(2.2)
$$
where $\epsilon_{ijk}$ is the fully antisymmetric third rank tensor.
For any two functionals $F$, $G$ we then have
$$
\{F,G \}= -\int_0^{\infty} \epsilon_{ijk}S_k \frac{\delta
F}{\delta S_i(x)}\frac{\delta G}{\delta S_j(x)}\: dx. \eqno(2.3)
$$

On taking into account (2.2)-(2.3), one can represent the equation
(2.1) in the following Hamiltonian form:

$$
{\bf S}_t=\frac{1}{x}\{H, {\bf S}\},  \eqno(2.4)
$$
where the Hamiltonian $H$ is given by

$$
H= \frac{1}{2} \int_0^{\infty} x {\bf S}_x^2\:dx. \eqno(2.5)
$$

Let us now define a new dependent complex-valued variable,

$$
w(x, t)=\frac{S_1+iS_2}{1-S_3}, \eqno(2.6)
$$
which is, at each fixed moment of time $t$, the stereographical
projection of the unit sphere onto the complex plane, $w:\:{\mathbb
S}^2 \to {\mathbb C}\cup \{\infty \}$.

In terms of this variable the equation (2.1) can be rewritten as

$$
iw_t=w_{xx}-2\frac{w_x^2{\bar w}}{1+|w|^2}+\frac{1}{x}w_x,
\eqno(2.7)
$$
and the Poisson brackets corresponding  to (2.2) take the
form{\footnote{ In the derivation of (2.8) we use the relations $
\{S_{\pm}(x), S_{\pm}(y)\}=0, \:\{S_{+}(x),
S_3(y)\}=-iS_{+}(x)\delta(x-y),\:\{S_{+}(x),
S_{-}(y)\}=2iS_3(x)\delta(x-y),$ where $S_{\pm}=S_1 \pm iS_2$, and
the Leibnits's rule.}}
$$
\{w(x), w(y)\}=\{{\bar w}(x), {\bar w}(y)\}=0,\:\:\{w(x), \bar {
w}(y)\}=-\frac{i}{2}(1+|w|^2)^2\delta(x-y). \eqno(2.8)
$$
The bracket (2.3) then becomes

$$
\{F, G \}= -\frac {i}{2} \int dx (1+|w(x)|^2)^2 \Bigl[\frac{\delta
F}{\delta w(x)} \frac{\delta G}{\delta {\bar w}(x)}-\frac{\delta
F}{\delta {\bar w}(x)} \frac{\delta G}{\delta w(x)} \Bigr ],
\eqno(2.9)
$$
and the evolution of the system will be described by the equation

$$
iw_t=-\frac{1}{2x}(1+|w|^2)^2\frac{\delta H}{\delta {\bar w(x)}} ,
\eqno(2.10)
$$
with the Hamiltonian

$$
H=2\int_0^{\infty} x\frac{w_x{\bar w}_x}{(1+|w|^2)^2}\: dx.
\eqno(2.11)
$$

It should be noticed that the following "complex extension"\enskip
of the system (2.1) is of interest of its own {\footnote {In
absence of the nonlinear component the second equation in (2.12)
can be interpreted as the free Shr\"odinger equation with an
effective mass. It is evident then that the first equation can be
obtained from the second by complex conjugation.}},
$$
ir_t=r_{xx}-\frac{2r_x^2s}{1+rs}+\frac{1}{x}r_x,\:\:\:\:\:
is_t=-s_{xx}+\frac{2s_x^2r}{1+rs}-\frac{1}{x}s_x.   \eqno(2.12)
$$
From (2.8) we obtain the Poisson brackets  of variables $r$ и $s$ in
the form

$$
\{r(x), s(y) \}=-i(1+rs)^2\delta (x-y),\:\:\:\{{\bar r}(x), {\bar
s}(y) \}=i(1+{\bar r}{\bar s})^2\delta(x-y). \eqno(2.13)
$$
The system (2.12), as well as equation (2.7), is completely
integrable (see Appendix) and have a Hamiltonian structure with the
Hamiltonian

$$
H=\int_0^{\infty} x\frac{r_xs_x}{(1+rs)^2}\: dx \eqno(2.14)
$$
and the equations of motion

$$
r_t=\frac{1}{x}\{H, r \},\:\:\:\:s_t=-\frac{1}{x} \{H, s \},
\eqno(2.15)
$$
and can be considered a model of the system of two coupled deformed
Heisenberg's magnets.

The Poisson brackets (2.13) can be found from the expression for
symplectic two-form,

$$
\Phi=i\int_0^{\infty} \Bigl [\frac{dr\wedge
ds}{(1+rs)^2}-\frac{d{\bar r}\wedge d{\bar s}}{(1+{\bar r}{\bar
s})^2)}\Bigr ]dx,\:\:\:\:\Phi=d\varphi,  \eqno(2.16)
$$
where

$$
\varphi=-i\int_0^{\infty} \Bigl [\frac{ds} {s(1+rs)}-\frac{d{\bar
s}}{{\bar s}(1+{\bar r}{\bar s}))}\Bigr ]dx, \eqno(2.17)
$$
thus (2.16) and (2.17) agree with the corresponding expressions
obtained in [18] for the standard Heisenberg magnet.

Notice also that the equation (2.7) is a bi-Hamiltonian system:
$$
i\left(\matrix {w \cr {\bar w} \cr }\right)_t =
G_1\left(\matrix{\frac{\delta H_1}{\delta w}\cr \frac{\delta
H_1}{\delta {\bar w}}\cr}\right)= G_2 \left(\matrix{\frac{\delta
H_2}{\delta {w}}\cr \frac{\delta H_2}{\delta {\bar w}}\cr}\right),
\eqno(2.18)
$$
where $H_1$ coincides with $H$ given by (2.11), the second
Hamiltonian $H_2$ reads as

$$
H_2=-i\int_0^{\infty} x \frac{w_x{\bar w}-{\bar
w}_xw}{(1+|w|^2)|w|^2}\:dx, \eqno(2.19)
$$
and $G_1=G_1(w, {\bar w}),\:G_2=G_2(w, {\bar w})$ are the so-called
Hamiltonian operators of the form

$$ G_1=\frac{1}{x(1+|w|^2)^2}\left(\matrix{0 & -1\cr -1  &\:\: 0
\cr}\right).  \eqno(2.20) $$

An expression for the matrix operator $G_2$ can be obtained from
results in paper [19] on the standard Heisenberg magnet but is too
cumbersome to be written here. Let us just mention that its matrix
entries contain a differential and an integral operator thus
rendering it non-local.

The relations (2.18)-(2.20) mean that the recursion operator of
the equation (2.7) under the assumption that ${\mathrm
det}\:G_2\neq 0$ is represented in the form

$$ R=G_1G_2^{-1}.                   \eqno(2.21) $$ Since (2.7) is
a completely integrable system, it admits infinitely many
integrals of motion, $\{I_n \}_{n=1}^{\infty}$ [7], in involution,
that is, satisfying $\{I_j, I_k \}=0$. In turn, this allows to
obtain hierarchies of the Poisson structures,

$$ I_n=R I_{n-1},             \eqno(2.22) $$ and the higher
equations of the deformed Heisenberg magnet ($j=0,1,...;t_0=t$),

$$
iw_{{t_j}}=R^jG_2\frac{\delta H_2}{\delta {\bar w}}.  \eqno(2.23)
$$

b). {\em Stability of certain solutions of equation (2.7).}

The problem of stability of stationary solutions of the equation
(2.7) is of interest since the equation contains the independent
variable $x$ explicitly. To analyze it, let $w=w_{st}+{\tilde w}$.
On linearizing (2.7), first on the trivial background $w_{st}=0$,
which corresponds, in terms of the magnetization vector, to the
vector ${\bf S}=(0,0,1)$, we obtain:

$$
i\tilde {w}_t(x,t)={\tilde w}_{xx}(x,t)+\frac{1}{x}{\tilde
w}_x(x,t).\eqno(2.24)
$$

Suppose that ($x > 0$)

$$
{\tilde w}(x, 0)={\tilde w}_0(x),\:\:\:\:\:{\tilde w}(0,t)={\tilde
w}_1(t).\eqno(2.25)
$$
Then the equation (2.24) can be solved by the Laplace
transformation in the $t$ variable under the additional assumption
that $|{\tilde w}(x,t)|< Me^{s_0t}$ with an $\:M > 0$ and $s_0
\geq 0$. Solving the arising equation and performing the inverse
transformation we find:

$$
{\bar w}(x, t)=\frac{1}{2\pi
i}\int_{a-i\infty}^{a+i\infty}e^{p\:t}
[C_0(x,p))J_0(\sqrt{-ip}\:x)]\:dp,    \eqno(2.26)
$$
where $J_0(.)$ is the Bessel function,

$$ C_0(x,p)=-i\int_0^x e^{-\int_0^xQ(\xi)\:d\xi}\:[\int_0^x
\frac{{\tilde
w}_0(y)}{J_0(\sqrt{-ip}\:y)}\:e^{\int_0^yQ(s)\:ds}\:dy\:]\:dx,
\eqno(2.27)
$$
$$ Q(x)=-2\sqrt{-ip}\:(\ln
J_0(\sqrt{-ip}\:x))_x+\frac{1}{x}, $$ ${\mathrm Re}\: a > 0$, the
path of integration is any straight line ${\mathrm Re}\: p=a
> s_0
> 0$, and the integral in (2.26) is understood in the sense of the principal
value. It is not difficult to see that the logarithmic divergencies
arising in the exponentials when integrating at the lower limit in
(2.28), cancel each other.

It follows from (2.26) that for a fixed $x$ the function $|{\tilde
w}(x, t)|$ grows with the $t$ increase, and, as in [7], we obtain
that the solution is unstable{\footnote {Of course, the stability
of that linearized "non-autonomous"\enskip equation is meant.}}:
an arbitrary localized initial perturbation of the system can grow
indefinitely as the time passes.

We now proceed to analyze stability of the stationary state
$w_{st}=ie^{i\theta(x)}$ where $\theta(x)=\ln
(x)+\theta_0,\:\theta_0 \in {\mathbb R}$ is a constant. This
solution is an example of a spiral-logarithmic structure found in
[10]: ${\bf S}=(\sin \theta,\:\cos \theta,\:0)$ {\footnote {Using
(2.11), it is easy to check that the Hamiltonian logarithmically
diverges on this solution in both limits and, thus, requires a
regularization.}}. On linearizing the equation (2.26) on this
background, we have,

$$ i{\tilde w}_t(x, t)={\tilde w}_{xx}(x, t)+\frac{1}{x}{\tilde
w}_x-i\frac{e^{-i\theta(x)}}{x^2}.  \eqno(2.29) $$
This equation
only differs from (2.24) by the presence of a non-homogeneous
term. Hence, its general solution is a sum of (2.26) and a partial
solution. It follows that it will be unstable as well.

Notice then, that the equation (2.7) admits a solution periodic in
$t$ of the form $w(x, t)=W(x)e^{ikt}$ with $k$ a real constant,
provided that the equation{\footnote {Removing the nonlinear term
we obtain here the stationary Schr\"odinger equation with the
Coulomb potential and an effective mass.}}

$$ W_{xx}-\frac{2W_x^2{\bar W}}{1+|W|^2}+\frac{1}{x}W_x+kW=0
\eqno(2.30) $$ has a solution. This suggests that the study of the
linearized stability is insufficient. The analysis of the
nonlinear stability requires more subtle methods [see, e. g. [20]
and literature cited therein].

\vskip2.4cm \hskip3.5cm  {\bf 3. LANDAU-LIFSHITS MAGNET}
 \vskip0.4cm

a). {\em {Canonical variables.}}

The fully anisotropic model of Landau-Lifshits has the form
{\footnote{It is well-known [21], that this model is one of the most
general completely integrable models admitting $2\times 2$ -matrix
Lax representations.}}

$$
{\bf S}_t={\bf S}\wedge {\bf S}_{xx}+{\bf S}\wedge J{\bf
S},\eqno(3.1)
$$
where $J={\mathrm diag}(J_1, J_2, J_3)$ are diagonal $3\times 3$
matrices, and $J_1,\:J_2,\:J_3$ are parameters of the anisotropy,
$J_1 < J_2 < J_3$.

The Hamiltonian for (3.1) can be written in the form,

$$
H=\frac{1}{2}\int_{-\infty}^{\infty}({\bf S}_x^2-{\bf S}J{\bf
S})dx, \eqno(3.2)
$$
or, using the variable $w$ defined in (2.6), as {\footnote {We
assume here that $w$ is a slowly decreasing function. In the case of
a decreasing $w$ one should add $J_3=4\beta $ to the density of the
Hamiltonian.}}

$$
H=\int_{-\infty}^{\infty}\Bigl (\frac{2(|w|_x^2+\alpha(w^2+{\bar
w}^2)-\gamma|w|^2)}{(1+|w|^2)^2}-\beta\Bigr) dx,  \eqno(3.3)
$$
where

$$
\alpha=\frac{J_2-J_1}{4},\:\:\beta=\frac{J_3}{4},\:\:\gamma= J_3-
\frac{J_1+J_2}{2}.   \eqno(3.4)
$$
Taking into account (2.9), from this we obtain the following
equation of motion for Landau-Lifshits magnet model
{\footnote{Notice that, as well as in the case of the deformed
Heisenberg magnet, we are able to obtain the corresponding complex
extension (see, [18]); we are not going to dwell on that here.}},

$$
iw_t=i\{H,w \}= -\frac{1}{2}(1+|w|^2)^2\frac{\delta H}{\delta
{\bar w}}, \eqno(3.5)
$$
or {\footnote {On taking the complex conjugated equation and
neglecting the nonlinear component, one can obtain the nonstationary
Shr\"odinger equation with the potential $V=-\gamma={\mathrm
const}$.}}
$$
iw_t=w_{xx}-2\frac{{\bar w}(w_x^2+\alpha)-\alpha w^3-\gamma
w}{1+|w|^2}-\gamma w . \eqno(3.6)
$$

Let us consider an implication of this form of the equation. Obvious
transformations lead to the following relation which contains the
parameter $\alpha$ only,

$$
i(|w|^2)_t=(w_x{\bar w}-w{\bar w}_x)_x+2\frac {w^2{{\bar
w}^2}_x-{\bar w}^2{w_x}^2}{1+|w|^2}+2\alpha(w^2-{\bar w}^2).
\eqno(3.7)
$$
Letting $w=\rho e^{i\varphi}$, где $\rho=\rho(x,
t),\:\varphi=\varphi(x, t),\:\rho,\:\varphi \in {\mathbb R}$, we
obtain:

$$
(\rho^2)_t=2(\rho^2\phi_x)_x-\frac{8\rho^3\rho_x\phi_x}{1+\rho^2}+4\alpha\rho^2\sin
2\phi .     \eqno(3.8)
$$

Defining the variables $R=\rho^2$ и $Q=2\rho^2\phi_x$, we now find
the following "conservation law":

$$
R_t=Q_x-2Q[\ln (1+R)]_x+4\alpha R\sin
(\int_{-\infty}^x\frac{Q}{R})\:dx.  \eqno(3.9)
$$
It is especially simple when $\alpha=0$, which corresponds to the
anisotropy of "the easy plan" type.

In a way similar to (2.18), one can produce a bi-Hamiltonian
structure  for (3.6) with $H_1 $ equal to $H$ defined by (3.3), the
Hamiltonian

$$
H_2=\int_{-\infty}^{\infty} \frac{w_x{\bar w}-{\bar
w}_xw}{(1+|w|^2)|w|^2}\: dx,  \eqno(3.10)
$$
and the Hamiltonian operators

$$
G_1=\frac{1}{(1+|w|^2)^2}\left(\matrix{0 & -1\cr -1  & 0
\cr}\right)      \eqno(3.11)
$$
and  $G_2$ being a matrix integro-differential operator [22]. In
terms of the variables $w$ и ${\bar w}$ the recursion operator can
be written as follows:

$$
R=G_1G_2^{-1} .    \eqno(3.12)
$$
It produces an hierarchy of the Poisson structures similar to (2.22)
and higher Landau-Lifshits equations similar to (3.5).

 b). {\em {The Dispersion relation. Stationary Landau-Lifshits equation.}}

Linearizing the equation complex conjugate to (3.6) and choosing
${\bar w}={\bar w}(x, t)$ in the form ${\bar w} \sim
\exp\{i(kx-\omega t)\}$, we have, $$ \omega=k^2-\gamma,
\eqno(3.13) $$ which gives a dispersion relation for the
Landau-Lifshits equation which is typical for magnets with an
exchange interaction [23]. In our case the group and phase
velocities are given by $v_{g}=\partial {\omega}/{\partial k}=2k$,
$v_{ph}=\omega/k=k-\gamma/k$, respectively (the latter is infinite
for $k=0$), implying that there is a dispersion in the system. The
propagation of a magnetization wave in this model is possible
under the condition $k^2 > \gamma=J_3-(J_1+J_2)/2 > 0.$

Letting $w=w(x-\mu t)=w(\xi)$ in (3.6) , where $\mu={\mathrm
const}$ is the velocity of a stationary profile wave, we obtain
the equation{\footnote {Stationary equations of another form for
the Landau-Lifshits hierarchy were considered from the viewpoint
of the Lie-algebraic approach in [24].}}

$$
w_{\xi \xi}+i\mu w_{\xi}-2\frac{{\bar w}(w_{\xi}^2+\alpha)-\alpha
w^3-\gamma w}{1+|w|^2}-\gamma w=0. \eqno(3.14)
$$
From this it is not difficult to obtain that

$$
(w_{\xi}{\bar w}-{\bar w}_{\xi}w)_{\xi}+i\mu\:(|w|^2)_{\xi}-
2\frac{w_{\xi}^2{\bar w}^2-{\bar
w}_{\xi}^2w^2}{1+|w|^2}+2\alpha(w^2-{\bar w}^2)=0.     \eqno(3.15)
$$
Letting $w(\xi)=\rho\:e^{i\phi}$, where
$\rho=\rho(\xi),\:\phi=\phi(\xi),\:\rho \in {\mathbb R}_{+},\:\phi
\in {\mathbb R}$, we then have:

$$
2(\rho^2)_{\xi}\phi_{\xi}+2\rho^2\phi_{\xi\xi}+\mu(\rho^2)_{\xi}
-8\frac{\rho^3\rho_{\xi}\:\phi_{\xi}}{1+\rho^2}+4\alpha \sin
2\phi=0.      \eqno(3.16)
$$
Let $\mu \neq 0$. Then, obviously, $\rho={\mathrm const},\:\phi=\pi
n/2,\:n=0,\:\pm 1,\:\pm 2,\:\pm 3$, satisfy (3.16). For
$\phi=\phi_0={\mathrm const}$ we obtain,
$\rho^2=\rho_0^2-(4\alpha/\mu)\sin(2\phi_0) \xi$, where
$\rho_0={\mathrm const}$; при $\rho={\tilde \rho}_0={\mathrm const}$
(3.16) is reduced to the equation of the pendulum:
$\phi_{\xi\xi}+(2\alpha/{\tilde \rho}_0)\sin(2\phi)=0$ (the
existence of other solutions remains an open problem).

Let now $\mu=0$, then from (3.16) it follows that

$$
\frac{({\rho^2})_{\xi}}{\rho^2}-\frac{4\rho\rho_{\xi}}{1+\rho^2}=C_1,
\:\:\:\:\:\:\:\phi_{\xi\xi}+C_1\phi_{\xi}+2\alpha\sin 2\phi=0,
\eqno(3.17)
$$
where $C_1$ is arbitrary constant. The first of these equations can
easily be integrated:

$$
\rho_{1,2}(\xi) = \frac{1}{2}(1\pm
\sqrt{1-4e^{-2(C_1\xi+C_2)}}\:)\:e^{C_1\xi+C_2},  \eqno(3.18)
$$
where $C_2$ is another arbitrary constant (we assume that
$e^{-2(C_1\xi+C_2)}< 1/4)$, and the second equation, which coincides
with the one of the pendulum with the friction {\footnote{In the
partial case $C_1=0$ this equation, can obviously be integrated in
terms of the elliptic functions.}}, admits, in particular, solutions
of the form $\phi=\pi n/2,\: n=0, \pm 1,\:\pm 2,\pm 3$.

Thus, solutions of the stationary Landau - Lifshits equation have
fairly non-trivial structure in the generic (fully anisotropic)
case. Their further study could bring a solution to an important
problem in the theory of dynamical systems - that of construction
of the phase graph for the equations (3.14) and
(3.6){\footnote{Phase graphs of the equation (3.1) in the case of
partial anisotropy have been studied in [25]. Phase graphs in the
fully anisotropic case have apparently not been considered yet.}}.
The same applies to the deformed Heisenberg magnet from the
previous section.

 \vskip2.4cm \hskip4.5cm  {\bf 4. ISHIMORI MAGNET}

\vskip0.4cm

a). {\em {Physical and geometrical interpretations.}}

The Ishimori magnet model in terms of the magnetization vector has
the form:

$$ {\bf S}_t={\bf S}\wedge({\bf S}_{xx}+\alpha^2{\bf
S}_{yy})+u_y{\bf S}_x+ u_x{\bf S}_y, \eqno(4.1) $$ $$
u_{xx}-\alpha^2u_{yy}=-2\alpha^2{\bf S} ({\bf S}_x \wedge {\bf
S}_y), \eqno(4.2) $$ where ${\bf S}(x,y,t)=(S_1, S_2, S_3)$ is a
three dimensional vector, $|{\bf S}|=1,\:u=u(x,y,t)$ is an
auxiliary scalar real-valued field, and the parameter $\alpha^2$
takes values $\pm 1$. The system is called the Ishimori-I magnet
(MI-I) in the case $\alpha^2=1$, the Ishimori-II magnet (MI-II) in
the case $\alpha^2=-1$. Mathematically, each of these cases
corresponds to different types of the equations (4.1) and (4.2).

The topological charge of the model (4.1)-(4.2), $$ Q_{T}=\frac
{1}{4\pi}\int_{{\mathbb R}^2}\int {\bf S}({\bf S}_x\wedge {\bf
S}_y)\:dx\: dy,  \eqno(4.3) $$ is invariant under the evolution of
the system. Since the homotopy group of the unit 2-sphere
$\pi_2(\tilde S^2)$ coincides with the group ${\mathbb Z}$ of
integers, the number $Q_T$ must be integer. According to (4.3),
the scalar function $u=u(x, y, t)$ is related to the density of
the topological charge production. The derivatives $u_x,\:u_y$ in
(4.1) play role of friction coefficients. Thus, (4.1) can be
interpreted as an equation of forced (by the friction power)
precession of the magnetization vector, and the system (4.1)-(4.2)
is self-consistent.

From the physical viewpoint, it is easy to see that there is a
non-local interaction in this system, on top of a local (exchange)
one. The mechanism of the former is unclear. Nevertheless, the
study of such systems is justified since stable localized
two-dimensional magnetic structures are observed in experiments.
An argument in favor of this assertion is the above-mentioned
gauge equivalence of the MI-II model and the DS-II model, which
describes quasi-monochromatic waves on the fluid surface [5], and
also a link found in [26] between the MI-I model and the nonlinear
Schr\"odinger equation with magnetic field.

Also helpful is another, hydrodynamical, interpretation of the
model (4.1)-(4.2). Namely, let $u_y=-v_1,\:u_x=v_2$, hence ${\bf
v}(x,y)=(v_1,\:v_2)$ is the velocity field of a fluid. Then the MI
model can be rewritten as follows:

$$ {\bf S}_t+v_1{\bf S}_x-v_2{\bf S}_y={\bf S}\wedge ({\bf
S}_{xx}+\alpha^2 {\bf S}_{yy}), $$ $$ \eqno(4.4) $$ $$
v_{2x}+\alpha^2v_{1y}=-2\alpha^2{\bf S}({\bf S}_x\wedge {\bf
S}_y). $$ If we define the stream function of the flow,
$v_1=-\chi_{1y},\:v_2=\chi_{1x}$, then the equation (4.2) with
$\alpha^2=-1$ (the MI-II model) implies the Poisson equation

$$ \chi_{1xx}+\chi_{1yy}=2{\bf S}({\bf S}_x\wedge {\bf S}_y),
\eqno(4.5) $$ that is, the stationary (the time $t$ is a parameter
here) vorticity equation with a source in the right hand side of
the magnitude proportional to the density of the topological
charge production (details on the equation of planar
hydrodynamical vortex can be found in [27]).

Let ${\tilde F}(x, y, t)= 2{\bf S}({\bf S}_x\wedge {\bf S}_y)$. On
taking one of the expressions of the form $\pm e^{\pm
\chi_1},\:e^{\chi_1}-e^{-2\chi_1},\:\pm \sinh {\chi_1},\:\pm \cosh
\chi_1,\:\pm \sin \chi_1,\:\pm \cos \chi_1$, for ${\tilde F}(x, y,
t)$, we obtain a closed completely integrable equation of elliptic
type for the function $\chi_1$. The solution of an appropriate
boundary-value problem for this equation must satisfy the
additional condition

$$ \frac{1}{8\pi} \int_{\mathbb R^2}\int \triangle \chi_1(x, y)
dx\:dy=N_0,\:\:\:N_0 \in {\mathbb Z}, \eqno(4.6a) $$ or
($r=\sqrt{x^2+y^2}$)

$$
\lim_{r \to \infty}\frac{1}{8\pi}\oint
(\chi_{1x}dy-\chi_{1y}dx)=N_0. \eqno(4.6b)
$$

b). {\em {New canonical variables.}}

Let us now consider another canonical variables. First, we pass from
the variable ${\bf S}$ to new variables $p$ и $q$ ($p,\:q \in
\mathbb R$) in (4.1)-(4.2), setting [28]:
$$
S_3(x, y, t)=p(x, y, t),\;\:\:\:S_{+}(x, y, t)=\sqrt{1-p^2(x, y,
t)}\:e^{iq(x, y, t)}.  \eqno(4.7)
$$
Expressions for Poisson brackets of the quantities $p$ и $q$
follow directly from (2.2), on taking into account that the
problem is two-dimensional,

$$
\{p({\bf r}), q({\bf r}^{\prime})\}= \delta({\bf r}-{\bf
r}^{\prime}),\;\: \{p({\bf r}), p({\bf r}^{\prime})\}= \{q({\bf
r}), q({\bf r}^{\prime})\}= 0,\:\:\:{\bf r}=(x, y),   \eqno(4.8)
$$
and then for the any two functionals $F$ and $G$ one can obtain:

$$
\{F, G \}= \int_{{\mathbb R}^2} \int \bigl [\frac {\delta
F}{\delta p({\bf r})}\frac{\delta G}{\delta q(\bf r)}-\frac
{\delta F}{\delta q(\bf r)}\frac{\delta G}{\delta p(\bf r)}\bigl
]\:dx\:dy.  \eqno(4.9)
$$
In terms of this variables the MI model (4.1)-(4.2) can be rewritten
as a Hamiltonian system,

$$ q_t=\frac{\delta H}{\delta
p}=-\frac{p_{xx}+\alpha^2p_{yy}}{1-p^2}-\frac{p\:(p_x^2+\alpha^2p_y^2)}
{(1-p^2)^2}-p\:(q_x^2+\alpha^2q_y^2)+u_yq_x+u_xq_y, $$ $$
p_t=-\frac{\delta H}{\delta q}=
(1-p^2)(q_{xx}+\alpha^2q_{yy})-2p\:(p_xq_x+\alpha^2p_yq_y)+u_yp_x+
u_xp_y,    \eqno(4.10) $$ $$
u_{xx}-\alpha^2u_{yy}=-2\alpha^2(p_yq_x-p_xq_y), $$ and for the
topological charge we will have:
$$ Q_T=\frac{1}{4\pi}\int_{{\mathbb R}^2}\int
(p_yq_x-p_xq_y)\:dx\:dy. \eqno(4.11) $$ Here the Hamiltonian $H$ has
the form {\footnote {Paper [29] contains an expression for the
Hamiltonian of the so-called modified MI different from (4.1) by the
sign in the last but one term. Thus, the Hamiltonian for the model
(4.1)-(4.2) seems to have been obtained here for the first time,
both in the $q,\:p$ and $w,\:{\bar w}$ variables, the latter being
defined below. Also, in contrast with the modified model, it is easy
to see that it is impossible to define the Clebsch variables in our
case.}} :
$$ H=H_1+H_2,\:\:\:H_1=\frac{1}{2}\int_{{\mathbb R}^2}\int \bigl [
\frac{p_x^2+\alpha^2p_y^2}{1-p^2}+(1-p^2)(q_x^2+\alpha^2q_y^2)\bigr
]dxdy , $$ $$ \eqno(4.12) $$ $$ H_2=\frac{1}{4} \int_{{\mathbb
R}^2}\int[\alpha^2A^2+B^2] \:dx\:dy, $$ where
$A=u_x,\:B=-\alpha^2u_y $, so that
$A_x+B_y=2\alpha^2(p_xq_y-p_yq_x)$; in this case it can be take
place the conditions:

$$
\frac{\delta A}{\delta
p}=C\delta_y(y-y^{\prime})\delta(x-x^{\prime}),\:\:\: \frac{\delta
B}{\delta p}=D\delta(y-y^{\prime})\delta_x(x-x^{\prime}),
$$
$$
\eqno(4.13)
$$
$$
\frac{\delta A}{\delta
q}=E\delta_y(y-y^{\prime})\delta(x-x^{\prime}),\:\:\: \frac{\delta
B}{\delta q}=F\delta(y-y^{\prime})\delta_x(x-x^{\prime}),
$$
where $C,\:D,\:E,\:F$ are some functions. Letting $D=C,\: F=E$ and
taking into account (4.10), we obtain the following relations on
the functions $C=C(x,y,t)$ and $E=E(x,y,t)$ (the symbol $<,>$
refers to the scalar product in $\mathbb R^2$, and $T$ stands for
the transposition):

$$
<\nabla u,\:
(\frac{1}{2\alpha^2}q_y+C_y,\:\frac{1}{2\alpha^2}q_x-C_x)^T>=0,
$$
$$
\eqno(4.14)
$$
$$
<\nabla u,\:
(-\frac{1}{2\alpha^2}p_y+E_y,\:-\frac{1}{2\alpha^2}p_x-E_x)^T>=0,
$$
from this we find:

$$
C(x, y, t)=C_0(u(x,
y))+\frac{1}{2\alpha^2}\int_s(u_xq_y+u_yq_x)ds,
$$
$$
\eqno(4.15)
$$
$$
E(x, y, t)=E_0(u(x,
y))+\frac{1}{2\alpha^2}\int_s(u_xp_y+u_yp_x)ds,
$$
where $C_0,\:E_0$ are arbitrary functionals, and the integration
goes along the characteristic $s$ of the equations (4.14). Assuming
that $E_0=C_0$, we see that the functional $C_0$ must obey an
additional condition:

$$
\frac {\delta C_0}{\delta u}\:(\frac {\delta u}{\delta q}-\frac
{\delta u}{\delta p})=\frac{2}{\alpha^2}u_{xy}. \eqno(4.16)
$$
Since $u_{xy}\ne 0$ in the generic case, from this it follows that
one more condition is necessary: ${\delta C}_0/\delta u \ne 0$
(if, of course, at this $\delta u/\delta q \ne \delta u/\delta p
$).

Now let us pass to the variable $w$ in (4.1)-(4.2), defined in (2.6)
 (assuming that $w=w(x, y, t)$)
 {\footnote{The reflection
 $(w,\:{\bar w}) \to (p,\:q)$ can by given by relations $q=-\arctan(i(w-
 {\bar w})/(w+{\bar w})),\:p=(|w|^2-1)/(1+|w|^2).$}} :

$$
iw_t=w_{xx}+\alpha^2w_{yy}-2\frac{{\bar
w}(w_x^2+\alpha^2w_y^2)}{1+|w|^2}+i(u_xw_y+u_yw_x),
$$
$$
\eqno(4.17)
$$
$$
u_{xx}-\alpha^2u_{yy}=4i\alpha^2\frac{w_x{\bar w}_y-{\bar
w}_xw_y}{(1+|w|^2)^2}.
$$
Then for the topological charge we obtain

$$
Q_T=-\frac{i}{2\pi}\int_{{\mathbb R}^2}\int \frac{w_x{\bar
w}_y-{\bar w}_xw_y}{(1+|w|^2)^2}\:dx\:dy . \eqno(4.18)
$$

The non-vanishing of the Poisson bracket for the canonical variables
$w(x,y), {\bar w(x,y)}$ comes along as in (2.8):

$$
\{w(x, y), {\bar w}(x^{\prime},
y^{\prime})\}=-\frac{i}{2}(1+|w|^2)^2 \delta({\bf r}-{\bf r
}^{\prime}), \:\:\:\:\:{\bf r}=(x, y).   \eqno(4.19)
$$
This allows to rewrite (4.1)-(4.2) in a transparently Hamiltonian
form:

$$
iw_t=-\frac{1}{2}(1+|w|^2)^2\:\frac{\delta H}{\delta {\bar w}}.
\eqno(4.20)
$$
Here $H$ is the Hamiltonian of the form

$$
H=H_1+H_2,\:\:\:H_1=2\int_{{\mathbb R}^2}\int \frac{w_x{\bar
w}_x+\alpha^2w_y{\bar w}_y}{(1+|w|^2)^2}\:dx\:dy,
$$
$$
\eqno(4.21)
$$
$$
H_2=\frac{1}{4}\int_{{\mathbb R}^2}\int [\alpha^2
u_x^2+u_y^2]\:dx\:dy,
$$
and we assume in the course of the derivation of the equations for
the model that the following conditions, analogous to (4.13), are
satisfied:

$$
\frac {\delta u_x}{\delta {\bar
w}}=-\frac{4iw_x}{\alpha^2(1+|w|^2)^2}\:\delta (x-x^{\prime})\:
\delta(y-y^{\prime}),\:\:\:\frac {\delta u_y}{\delta {\bar
w}}=-\frac{4iw_y}{(1+|w|^2)^2}\:\delta
(x-x^{\prime})\:\delta(y-y^{\prime})  . \eqno(4.22)
$$

Clearly, all three representations of the MI model, (4.1)-(4.2),
(4.10) и (4.17) are equivalent.

Notice also that, one can define a "complex extension" of the
system (4.17) analogous to the ones above. Letting formally
$w_1={\bar w}$, one obtains,

$$
iw_t=w_{xx}+\alpha^2w_{yy}-2\frac{w_1(w_x^2+\alpha^2w_y^2)}{(1+ww_1)^2}+i(u_xw_y+u_yw_x),
$$
$$
iw_{1t}=-(w_{1xx}+\alpha^2w_{1yy})+2\frac{w(w_{1x}^2+\alpha^2w_{1y}^2)}{(1+ww_1)^2}
+i(u_xw_{1y}+u_yw_{1x}), \eqno(4.23)
$$
$$
u_{xx}-\alpha^2u_{yy}=4i\alpha^2\frac{w_xw_{1y}-w_{1x}w_y}{(1+ww_1)^2}.
$$

This system can be interpreted as a model of two coupled Ishimori
magnets. Nontrivial Poisson brackets follow from (4.19):

$$
\{w(x, y),  w_1(x^{\prime},
y^{\prime})\}=-\frac{i}{2}(1+ww_1)^2\delta({\bf r}-{\bf r
}^{\prime}),\:\:\: \{{\bar w}(x, y), {\bar w}_1(x^{\prime},
y^{\prime})\}=\frac{i}{2}(1+{\bar w}{\bar w}_1)^2\delta({\bf
r}-{\bf r }^{\prime}),       \eqno(4.24)
$$
and the "topological charge"\enskip of this model is {\footnote
{In general, $w_1 \ne {\bar w}$, thus $Q_T$ can be non-integer and
even complex. Such a situation, including an interpretation of the
quantity $Q_T$, should be considered separately.}}

$$
Q_T=-\frac{i}{2\pi}\int_{{\mathbb R}^2} \int
\frac{w_xw_{1y}-w_{1x}w_y}{(1+ww_1)^2}\:dx\:dy . \eqno(4.25)
$$
The equations of motion (4.23) are Hamiltonian:

$$
iw_t=-\frac{1}{2}(1+ww_1)^2\frac {\delta H}{\delta w_1}
,\:\:\:\:\:\: iw_{1t}=\frac{1}{2}(1+ww_1)^2\frac {\delta H}{\delta
w} , \eqno(4.26)
$$
where
$$
H=H_1+H_2,\:\:\:H_1=2\int_{{\mathbb R}^2} \int \frac{w_x
w_{1x}+\alpha^2w_y w_{1y}}{(1+ww_1)^2}\:dx\:dy,
$$
$$
\eqno(4.27)
$$
$$
H_2=\frac{1}{4}\int_{{\mathbb R}^2}\int [\alpha^2
u_x^2+u_y^2]\:dx\:dy ,
$$
and we suppose that

$$
\frac {\delta u_x}{\delta
w_1}=-\frac{4iw_y}{\alpha^2(1+ww_1)^2},\:\:\:\:\:\:\frac {\delta
u_y}{\delta w_1}=-\frac{4iw_x}{(1+ww_1)^2},
$$
$$
\eqno(4.27a)
$$
$$
\frac {\delta u_x}{\delta
w}=\frac{4iw_{1y}}{\alpha^2(1+ww_1)^2},\:\:\:\:\:\:\frac {\delta
u_y}{\delta w}=\frac{4iw_{1x}}{(1+ww_1)^2}.
$$

Returning to (4.17), we introduce the complex coordinates
$z=x+iy,\:{\bar z}=x-iy$, so that
$\partial_z=1/2(\partial_x-i\partial_y),\:\partial_{\bar
z}=1/2(\partial_x+i\partial_y),\:dx\:dy=(i/2)dz\wedge d{\bar z}$ and
rewrite (4.18) and (4.21) in terms of these variables.

i). Let $\alpha^2=1$, that is, the MI-I model is considered. In this
case we obtain:

$$
iw_t=4w_{z\bar z}-8{\frac{w_zw_{\bar z}}{1+|w|^2}}{\bar
w}-2(w_zu_z-w_{\bar z}u_{\bar z}),
$$
$$
\eqno(4.28)
$$
$$
u_{zz}+u_{\bar z\bar z}=4\frac{w_z{\bar w}_{\bar z}-w_{\bar
z}{\bar w}_z}{(1+|w|^2)^2}.
$$
The topological charge is given by
$$
Q_T=\frac{i}{2\pi}\int \int \frac{w_z{\bar w}_{\bar z}-w_{\bar
z}{\bar w}_z}{(1+|w|^2)^2}\:dz\wedge d{\bar z}, \eqno(4.29)
$$
and the Hamiltonian is
$$
H=2i\int \int \frac{w_z{\bar w}_{\bar z}+w_{\bar z}{\bar
w}_z}{(1+|w|^2)^2}\:dz\wedge d{\bar z}+\frac{i}{2}\int \int u_z
u_{\bar z}\:dz\wedge d{\bar z}. \eqno(4.30)
$$

ii). Let $\alpha^2=-1$. We then have the MI-II model,

$$
iw_t=2(w_{zz}+w_{{\bar z}{\bar z}})-4\frac{w_z^2+w_{\bar
z}^2}{1+|w|^2}\:{\bar w}-2(w_zu_z-w_{\bar z}u_{\bar z}) ,
$$
$$
\eqno(4.31)
$$
$$
u_{z{\bar z}}=-2\frac{w_z{\bar w}_{\bar z}-w_{\bar z}{\bar w}_z}
{(1+|w|^2)^2}.
$$
The expression for the topological charge coincides with (4.29), and
for the Hamiltonian we have:

$$
H=2i\int \int \frac{w_z{\bar w}_z+w_{\bar z}{\bar w}_{\bar
z}}{(1+|w|^2)^2}\:dz \wedge d{\bar z}-\frac{i}{4}\int
\int(u_z^2+u_{\bar z}^2)\:dz \wedge d{\bar z}.  \eqno(4.32)
$$

Notice also that the Hamiltonian of the MI-I magnet and its
topological charge are related, as follows from (4.29) and (4.30),
by the inequality of Bogomol'nyi, which is a lower estimate for the
Hamiltonian taking into account all the dynamical configurations.
Namely,

$$
H \geq 4\pi Q_{T}.  \eqno(4.33)
$$
Comparing (4.29) and (4.32) one can see that for the MI-II model
such an estimate does not exist.

 c). {\em Hamiltonians and topological charges for some of the simplest solutions.}
 {\footnote {
 Similar calculations were given in [29] for the case of the modified MI
 and certain other systems.}}

Equations (4.1)-(4.2) can be interpreted as the compatibility
 conditions for the following overdetermined matrix systems on the function
$\Psi=\Psi(x, y, t)$:

$$
\Psi_y=\frac {1}{\alpha}S\Psi_x  ,\eqno(4.34)
$$
$$
\Psi_t=-2iS\Psi_{xx}+Q\Psi_x  ,\eqno(4.35)
$$
where $Q=u_yI+\alpha^3u_xS+i\alpha S_yS-iS_x, \: \Psi=\Psi(x,y,t)\in
Mat(2,\mathbb{C}),  \:S=\sum_{i=1}^3 S_i\sigma_i,\: \sigma_i$ are
the standard Pauli matrices, $I$ is the unit $2\times 2$ matrix. By
the definition, the $S$ matrices have the properties:
$S=S^{\ast},\;S^2=I,\;\det S=-1,\:{\mathrm Sp}\:S=0$ (the asterisk
stands for the Hermitian conjugation).

For a future reference, let us provide an expression for the $S$
matrices in terms of the $w$ variable:

$$
S=\left(\matrix{\frac{|w|^1-1}{1+|w|^2}& \frac{2{\bar
w}}{1+|w|^2}\cr \frac{2w}{1+|w|^2}&-\frac{|w|^2-1}{1+|w|^2}}
\right),\eqno(4.36)
$$
and consider some of the simplest examples of calculations.

1. Let $\alpha^2=-1$ in (4.17), that is, the MI-II model is
considered. As was shown in [26], the conditions $$ w_{\bar
z}=0,\:\:\:\:\:\:w_z=0 , \eqno(4.37) $$ are then compatible with
(4.31). The first and second conditions here mean the presence of
instanton and anti-instanton sectors, respectively, in the MI-II
model. Consider the instanton sector assuming that
$w(z)=((z-z_0)/\lambda)^n$ [30]{\footnote {The choice of a more
general solution, say, in the form of the Belavin-Polyakov
instanton (linear-fractional function with complex poles) [2],
unfortunately, significantly complicates the calculations.}},
where $n \in {\mathbb Z}_{+},\:\lambda \in {\mathbb C}$ (the $z_0$
and $\lambda$ characterize, respectively, the position and size of
the instanton). A calculation by the relation (4.39) gives:

$$ Q_T= \frac{i}{2\pi}\int\int \frac{|w_z|^2}{(1+|w|^2)^2}\:dz
\wedge d{\bar z}=n , \eqno(4.38) $$ and $H_1=0$ by (4.32). To find
$H_2$, one has to know the function $u$. Using the second equation
in (4.31) with $w_{\bar z}=0$ and returning to the Cartesian
coordinates, we obtain ($z_0=x_0+iy_0$),

$$ \triangle
u=-8\lambda^{2n}\frac{[(x-x_0)^2+(y-y_0)^2]^n}{[\lambda^{2n}+[(x-x_0)^2+(y-y_0)^2]^n]^2
},                    \eqno(4.39) $$ which implies that
{\footnote{The following integral can be simplified by a change of
variables and subsequent contour integration, but the remaining
integral, apparently, cannot be calculated explicitly.}}

$$ u(x, y)=-8\lambda^{2n}\int_{{\mathbb R}^2}\int
G_0(x-x^{\prime},y-y^{\prime})\frac{[(x^{\prime}-x_0)^2+(y^{\prime}-y_0)^2]^n}
{[\lambda^{2n}+[(x^{\prime}-x_0)^2+(y^{\prime}-y_0)^2]^n]^2}\:
dx^{\prime}\:dy^{\prime},    \eqno(4.40) $$ where $G_0(x,
y)=(1/2\pi)\ln (x^2+y^2)$ is the Green function of the
two-dimensional Laplace operator. Thus, the energy of the
instanton solution on the formal level is given by
{\footnote{Obviously, the Hamiltonian is positive in the domain
where $|u_y| > |u_x|$.}}

$$
H=\frac{1}{4}\int_{{\mathbb R}^2}\int (u_y^2-u_x^2)\:dx\:dy.
\eqno(4.41)
$$

The whole instanton sector is then split into disjoint classes
each corresponding to the relevant value of the $Q_T$ quantity.

2. Let us consider the MI-I model ($\alpha^2=1$) and show that
instanton solutions exist in there as well{\footnote {This is not
surprising, albeit apparently went unnoticed in the literature,
given that in the "static limit"\enskip the MI-I model turns into
the elliptic version of the nonlinear $O(3)$ $ \sigma $-model for
which the instanton solutions were constructed initially. Notice
also that the model was solved by the inverse scattering method in
[31]-[32].}},{\footnote{From the viewpoint of the
higher-dimensional inverse scattering method and the dressing
procedures for solutions, the characteristic variables
$\xi=(y-x)/2$ and $\eta=(x+y)/2$ [14], [15] are more natural than
$z$ and $\bar z$.}}. Indeed, for $w_{\bar z}=0$ the system (4.28)
is reduced to the following one,

$$ w_t=-2u_zw_z,\:\:\:\:\:u_{zz}+u_{{\bar z}{\bar
z}}=4\frac{w_z{\bar w}_{\bar z}}{(1+|w|^2)^2} . \eqno(4.42) $$
Differentiating the first relation in $\bar z$, we obtain that
$u_{z\bar z}=0$, whence the compatibility is achieved if

$$ u_{xx}=4\frac{w_z{\bar w}_{\bar z}}{(1+|w|^2)^2} , \eqno(4.43)
$$ or

$$ u(x, y, t)=4\int_{-\infty}^x dx^{\prime}
\int_{-\infty}^{x^{\prime}}dx^{\prime \prime}
\frac{|w_z|^2}{(1+|w|^2)^2}+f_0(y,t)x+c_1, \eqno(4.44) $$ where
$c_1$ is an arbitrary constant, and $f_0(.,.)$ is an arbitrary
function. For the instanton solution $w(z)$ of the same form as in
the previous case the number $Q_T=0$ and the function

$$ u(x, y, t)= \frac{4n^2}{\lambda^{2n}}\int_{-\infty}^x
dx^{\prime}\int_{-\infty}^{x^{\prime}}dx^{\prime \prime}
\frac{[(x^{\prime
\prime}-x_0)^2+(y-y_0)^2]^{n-1}}{[\lambda^{2n}+(|x^{\prime
\prime}-x_0|^2+|y-y_0|^2)^n]^2}+f_0(y,t)x+c_1. \eqno(4.45) $$ The
expression for the Hamiltonian takes the form,

$$
H=4\pi n+\frac{1}{4}\int_{{\mathbb R}^2} \int (u_x^2+u_y^2)\:dx\:
dy.           \eqno(4.46)
$$

3. Let us calculate the topological charge and the Hamiltonian for
a solution of the form of a spiral structure. Namely, let ${\bf
S}=(0,\:\sin \Phi_1,\:\cos \Phi_1)$, where $\Phi_1=\delta_0t+
\alpha_0x+\beta_0y+\gamma_0,\:\alpha_0,\:\beta_0,\:\gamma_0,\:\delta_0
\in \mathbb {R}$ are parameters, that is, the solution is a
two-dimensional spiral structure [16]; then, according to (2.6)
(see also (4.36)), $w(z, {\bar z})=i\tan (\Phi_1/2),\:
\Phi_1=\delta_0t+\alpha z+{\bar \alpha}{\bar
z}+\gamma_0,\:\alpha=\alpha_0/2+\beta_0/(2i)$. It follows from
(4.29) that $Q_T=0$.

To determine the function $u=u(x,\:y,\:t)$, one has to substitute
the function $w(z, \bar z)$) in the equations (4.28), (4.31),
which gives two linear equations for $u$. Assuming their
compatibility and integrating, we find ($\alpha^2=-1$)[16]:

$$ u(x,y)=g_0(y+\frac{\beta_0}{\alpha_0}x)+\int_s g_1
(y(s^{\prime})+\frac{\beta_0}{\alpha_0}x(s^{\prime}),t)\:ds^{\prime},
\eqno(4.47) $$ where $g_0,\:g_1$ are arbitrary functions such that
$g_0$ is constant on the characteristic
$y+(\beta_0/\alpha_0)x={\mathrm const}$, and $s$ is the
characteristic taken to be the integration path.

Similarly, for $\alpha^2=1$ we have:

$$
u(x, y)=g_2(y-\frac{\beta_0}{\alpha_0}x)+\int_{s_1} g_3
(y(s^{\prime})-\frac{\beta_0}{\alpha_0}x(s^{\prime}),t)\:ds^{\prime},
\eqno(4.48)
$$
where $g_2,\:g_3$ are arbitrary functions, and $g_2$ is constant on
the characteristic $s_1$ $y-(\beta_0/\alpha_0)x={\mathrm const}$.

The substitution $w=w(z,\:{\bar z})$ in (4.30) and (4.32) (in both
MI-I and MI-II cases) leads to divergence of the Hamiltonian
$H_1$, and, therefore, that of the Hamiltonian $H$ as a whole,
since the functional $H_2$ is finite.

4. As was first shown in [33] (see also [14], [16]), in the
reflectionless section of the MI-II model the system (4.34)-(4.35)
can be written in the form

$$ \tilde \Psi_{\bar z}=0, \:\:\:\tilde {\Psi}_t+2i\tilde
\Psi_{zz}=0. \eqno(4.49) $$ In turn, the latter system has
well-known polynomial solutions describing vortex states, ($\tilde
\Psi=\{\tilde \Psi_{ij}\},i,j=1,2,\:{\tilde \Psi}_{22}=\bar
{\tilde \Psi}_{11},\:{\tilde \Psi}_{12}=-\bar {\tilde
\Psi}_{21}$)[33]: $$ \tilde
\Psi_{11}(z,t)=\sum_{j=0}^{N_1}\sum_{m+2n=j}\frac{a_j}{m!
n!}(-\frac{1}{2} z)^m(-\frac{1}{2} it)^n, $$ $$ \eqno(4.50) $$ $$
 \tilde
\Psi_{21}(z,t)=\sum_{j=0}^{M_1}\sum_{m+2n=j}\frac{b_j}{m!
n!}(-\frac{1}{2} z)^m(-\frac{1}{2} it)^n, $$ where $N_1$ is an
integer, $M_1=N_1-1,\:a_j,\:b_j$ are complex numbers, and the
inner summations run over all $m,\:n \geq 0$ such that $m+2n=j$.
In particular, in this simplest case $N_1=1$ it follows that

$$
\tilde {\Psi}_{11}= a_0+a_1z,\:\:{\tilde \Psi}_{21}=b_0.
\eqno(4.51)
$$

Let us employ now a dressing (say, the Darboux dressing [16])
relation for the matrix S {\footnote {Its structure is identical,
at least, for all the models of magnets treated here and all known
methods of their solutions.}}, ${\tilde S} = {\tilde
\Psi}S^{(1)}{\tilde \Psi}^{-1}$, where $S^{(1)}$ is the initial
solution of the system (4.1)-(4.2), assuming that
$S^{(1)}=\sigma_3$. This leads to the so-called "one-lump"\enskip
stationary solution, which we write here in terms of the
stereographic projection,

$$
w(z, {\bar z})=\frac{{\bar b}_0}{{\bar a}_1({\bar z}-{\bar
z}_0)+{\bar d}_0}, \eqno(4.52)
$$
where $d_0=a_0-a_1z_0,\:z_0$ is the coordinate of the vortex center
on the complex plan.

It is known [33], that $ Q_T=1$ for such a solution. Let us
calculate the function $u=u(x, y)$. In the case under consideration
the second equation in (4.31) is reduced to the following,

$$
u_{z{\bar
z}}=\frac{2|a_1|^2|b_0|^2}{(|a_1(z-z_0)+d_0|^2+|b_0|^2)^2}.
\eqno(4.53)
$$
From this we find

$$
u(x, y)=2 |a_1|^2|b_0|^2\int_{{\mathbb R}^2}\int
G_0(x-x^{\prime},y-y^{\prime})\frac{1}{(|a_1(z^{\prime}-z_0)+d_0|^2+|b_0|^2)^2}\:
dx^{\prime}\:dy^{\prime}.          \eqno(4.54)
$$

Taking into account that $H_1=0$, we now obtain from (4.32) the
Hamiltonian of the one-lump solution in the form

$$ H= \int_{{\mathbb R}^2}\int (u_y^2-u_x^2)\:dx\:dy. \eqno(4.55)
$$ The left hand side is positive in the planar domain where
$|u_y| > |u_x|$ in the same way as in (4.41).

\vskip2.4cm \hfil{{5. \bf CONCLUSION}}\hfil

\vskip0.8cm

The results for the Ishimori magnet show, in particular, that the
Hamiltonians and topological charges cannot always be calculated
analytically even for the simplest solutions, and numerics are
required for specific Cauchy problems. In this respect, it is
especially interesting, in our opinion, to check the hypothesis
[16] of possibility of a phase transition in the model which
involves a change of topology and symmetry properties of the
system.

Concerning the "extended" \enskip systems (2.12) and (4.23), we
would like to point out that if the initial systems are gauge
equivalent to the nonlinear Schr\"odinger equation with an
integral nonlinearity [8,9] (the Davy-Stewartson-II model [5] in
the IM-II case), then it is interesting and important to find
objects gauge equivalent to the extended systems.

Overall, the representations considered in this paper can,
hopefully, be useful in studying other (1+1)--- (and more
realistic (2+1)--- and (3+1)---) dimensional models of magnets, $
\sigma $-models and chiral fields, including nonintegrable cases.

The author is indebted to P. Kulish for support.

\vskip0.8cm \hfil {\bf {APPENDIX}}\hfil \vskip0.8cm

We provide here an expression for a Lax pair for the system (2.12).
First, define the function $D=D(x, t) \in Mat(2, {\mathbb C}),\:x
>0$, of the form

$$
D(x, t)= \left(\matrix{\frac{sr-1}{1+sr}& \frac{2s}{1+sr}\cr
\frac{2r}{1+sr}&-\frac{sr-1}{1+sr}} \right).    \eqno(A.1)
$$
The matrix $D$ has the following properties: $D^2=I,\:{\mathrm
Sp}\:\:D = 0,\:{\mathrm det}\:D =-1$. However, unlike the matrix
$S$, it is not Hermitian. A straightforward calculation shows that
(2.12) is the compatibility condition for the following
overdetermined linear system of equations:

$$
\Psi_x=-\frac{i}{2}\lambda D\Psi,\:\:\:\:\Psi_t=(\frac{\lambda}{2}
D_xD+\frac{i}{2}\lambda^2D)\Psi, \eqno(A.2)
$$
where $\Psi=\Psi(x, t, \lambda)\in Mat(2, {\mathbb C}),\:\lambda =
\lambda(x, t, u),\:u \in {\mathbb C}$. This means that the parameter
$u$ plays the role of a "hidden" \enskip spectral parameter, so that
the conditions

$$
\lambda_x=\frac{\lambda}{x},\:\:\:\:\:\:\lambda_t=-\frac{2\lambda^2}{x},
\eqno(A.3)
$$
or

$$
\lambda=\frac{x}{2(t+u)} ,   \eqno(A.4)
$$
are fulfilled, and the matrices $D$ satisfy the equation

$$
iD_t=\frac{1}{2}[D, D_{xx}]-\frac{1}{x}D_xD. \eqno(A.5)
$$

Thus, we have a non-isospectral deformation of the associated linear
system (the case of a single deformed Heisenberg magnet was
considered in [8]). Notice also that so far the inverse scattering
transform method has not been applied  to such systems at the full
scale.

\vskip2.4cm

\hfil{\bf {REFERENCES}}\hfil \vskip0.6cm

1. \parbox[t]{12.7cm}
       {{\em L.D.Landau,} - Collected Papers, (1969),
       128-143, Moscow, Nauka.}
\vskip0.3cm

2. \parbox[t]{12.7cm}
       {{\em A.A.Belavin and A.M.Polyakov,} - JETP Letters,  22 (1975),
       503-506.}
\vskip0.3cm
3. \parbox[t]{12.7cm}
       {{\em A.M.Perelomov,} - Uspekhi Fizich. Nauk, 134 (1990), 577-609.}

\vskip0.3cm
4. \parbox[t]{12.7cm}
       {{\em Ju.M.Izumov and Ju.N.Skriabin,} - Statistical mechanics of the magnet-nonordered
       medium, (1987), Moscow, Nauka.}

\vskip0.3cm
 5. \parbox[t]{12.7cm}
     {{\em V.D.Lipovskii and A.V.Shirokov,}  -  Funct. An. and Appl, 23 (1989), 65.}

\vskip0.3cm
6.   \parbox[t]{12.7cm}
     {{\em O.I.Mokhov,} - Symplectic and Poisson geometry on the spaces of the loops
     of the smooth manifolds and integrable equations (2004), -
     Institute of the Computers' s researches, Moscow-Igevsk.}

\vskip0.3cm

7.   \parbox[t]{12.7cm}
  {{\em A.V.Mikhailov and A.I.Jaremchuk,} - JETP Letters, 36 (1982), 78.}

\vskip0.3cm

8.   \parbox[t]{12.7cm}
  {{\em S.P.Burtsev, V.E.Zakharov and A.V.Mikhailov,} - TMP, 70 (1987), 323.}

\vskip0.3cm

9.  \parbox[t]{12.7cm}
    {{\em K.Porsezian and M.Lakshmanan,} - J.Math.Phys., 32, (1991), 2923.}

\vskip0.3cm

10. \parbox[t]{12.7cm}
    {{\em E.Sh.Gutshabash,} - JETP Letters, 73 (2001), 317-319.}
\vskip0.3cm

11. \parbox[t]{12.7cm}
    {{\em E.K.Sklyanin,} - Preprint No E-3-79 LOMI (1979), Leningrad.}
\vskip0.3cm

12. \parbox[t]{12.7cm}
    {{\em A.I.Bobenko,} - Zapiski Nauchnych Seminarov POMI, 123, (1983), 58.}
\vskip0.3cm

13. \parbox[t]{12.7cm}
    {{\em Niam-Ning Huang, Hao Cai and ot.,} - arxiv:nlin.SI/0509042.}
\vskip0.3cm

14. \parbox[t]{12.7cm}
    {{\em B.G.Konopelchenko,} - Solitons in Multidimensions, (1993), World Scientific.}

\vskip0.3cm

15. \parbox[t]{12.7cm}
    {{\em E.Sh.Gutshabash,} - Zapiski Nauchnych Seminarov POMI, 17 (2002), 155-168.}

\vskip0.3cm

16. \parbox[t]{12.7cm}
    {{\em E.Sh.Gutshabash,} - JETP Letters, 78 (2003), 1257.}

\vskip0.3cm

17. \parbox[t]{12.7cm}
    {{\em Qing Ding,} - arxiv.math.DG/0504288.}

 \vskip0.3cm

18. \parbox[t]{12.7cm}
     {{\em  A.N.Leznov and A.V.Razumov,} - J.Math.Phys., 35 (1994), 1738.}

\vskip0.3cm
19. \parbox[t]{12.7cm}
    {{\em E.V.Ferapontov,} - TMP, 91 (1992), 452.}

\vskip0.3cm
20. \parbox[t]{12.7cm}
    {{\em V.G.Mahan'kov, Ju.P.Rybakov and V.I.Sanuk,} - Uspekhi Fizich. Nauk, 162 (1994),
    121-148.}
\vskip0.3cm

21. \parbox[t]{12.7cm}
     {{\em L.A.Takhtadjan and L.D.Faddeev,} - Hamiltonian approach in theory of Solitons
     (1986), Moscow, Nauka.}

\vskip0.3cm
 22.  \parbox[t]{12.7cm}
     {{\em  E.Barouch, A.Fokas and V.Papageorgiou,} - J.Math.Phys., 29 (1988), 2628.}

\vskip0.3cm 23. \parbox[t]{12.7cm}
     {{\em E.M.Lifshits and L.P.Pitaevskii,} - Statistical Physics, part 2 (1978), Moscow,
     Nauka.}

\vskip0.3cm
 24. \parbox[t]{12.7cm}
     {{\em V.G.Bariakhtar, E.D.Belokolos and P.I.Golod,} - In: Modern Problems in
     Magnetism, (1986), 30, Kiev, Naukova dumka.}

\vskip0.3cm
25. \parbox[t]{12.7 cm}
    {{\em V.M.Eleonskii, N.N.Kirova and N.E.Kulagin,} - JETP, 79 (1980), 321.}

\vskip0.3cm
 26. \parbox[t]{12.7cm}
    {{\em V.G.Mikhalev,} - Zapiski Nauchnych Seminarov POMI, 189 (1991), 75.}

\vskip0.3cm
27. \parbox[t]{12.7cm}
    {{\em E.Sh.Gutshabash,} - Zapiski Nauchnych Seminarov POMI, 19, (2006), 119-133.}

\vskip0.3cm
28. \parbox[t]{12.7cm}
    {{\em H.C. Fogedby,} - J.Phys.A: Math.Gen., 13 (1980), 1467-1499.}

\vskip0.3cm
29. \parbox[t]{12.7cm}
    {{\em L.Martina, G.Profilo, G.Soliani and L.Solombrino,} - Phys. Rev.B, 49 (1994),
    12915.}

\vskip0.3cm
30. \parbox[t]{12.7cm}
     {{\em R.Rajaraman,} - Solitons and Instantons (1982), North-Holland
     Publishing Company, Amsterdam-New York-Oxford.}

\vskip0.3cm
31. \parbox[t]{12.7cm}
    {{\em E.Sh.Gutshabash and V.D.Lipovskii,} - TMP, 90 (1992), 175.}

\vskip0.3cm
32. \parbox[t]{12.7cm}
    {{\em G.G.Varzugin, E.Sh.Gutshabash and V.D.Lipovskii,} - TMP, 104 (1995), 513.}

\vskip0.3cm
33. \parbox[t]{12.7cm}
    {{\em Y.Ishimori,} - Progr.Teor.Phys, 72 (1984), 33.}

\end{document}